# Efficient Terahertz Harmonic Generation with Coherent Acceleration of Electrons in the Dirac Semimetal Cd$_3$As$_2$

Bing Cheng[1]†, Natsuki Kanda[2]†, Tatsuhiko N. Ikeda[2], Takuya Matsuda[2], Peiyu Xia[2], Timo Schumann[3], Susanne Stemmer[3], Jiro Itatani[2], N. P. Armitage[1], and Ryusuke Matsunaga[2,4]*

[1]*The Institute for Quantum Matter and Department of Physics and Astronomy, The Johns Hopkins University, Baltimore, MD 21218, USA.*

[2]*The Institute for Solid State Physics, The University of Tokyo, Kashiwa, Chiba 277-8581, Japan.*

[3]*Materials Department, University of California, Santa Barbara, California 93106-5050, USA.*

[4]*PRESTO, Japan Science and Technology Agency, 4-1-8 Honcho Kawaguchi, Saitama 332-0012, Japan.*

†The authors contributed equally.
*e-mail: matsunaga@issp.u-tokyo.ac.jp

**We report strong terahertz (~10$^{12}$ Hz) high harmonic generation in thin films of Cd$_3$As$_2$, a three-dimensional Dirac semimetal at room temperature. The third harmonics is detectable with tabletop light source and can be as strong as 100 V/cm by applying the fundamental field of 6.5 kV/cm inside the film, showing an unprecedented efficiency for terahertz frequency conversion. Our time-resolved terahertz spectroscopy and calculations also clarify the microscopic mechanism of the nonlinearity originating in the coherent acceleration of Dirac electrons in momentum space. Our results provide clear insights for nonlinear current of Dirac electrons driven by terahertz field under an influence of scattering, paving the way toward novel devices for high-speed electronics and photonics based on topological semimetals.**

One of the major advances in modern photonics is the realization of intense phase-stable light fields, which have revealed highly-intriguing nonlinear and nonperturbative light-matter interactions in atomic or molecular gases and condensed matter. A prominent example is high harmonic generation (HHG), *i.e.,* the production of coherent high-energy photons by multiplication of an incident photon energy, which has been developed in gaseous media for attosecond science [1] and has also been utilized for high-resolution angle-resolved photoemission spectroscopy (ARPES) [2]. More recently, HHG in semiconductors with mid-infrared (IR) excitation has been reported [3,4], opening a new route toward laser-based stable and compact soft X-ray sources.

Efficient third harmonic generation (THG) in the terahertz (THz) frequency regime has been also reported in a superconducting film resonating with a collective mode [5]. Realization of such strong THz nonlinearity at room temperature is highly desired for high-speed electronics and for frequency mixing in sensitive detection of far-infrared wave. From this perspective, graphene, a two-dimensional (2D) honeycomb carbon sheet, has attracted tremendous attention as a candidate for efficient THz frequency multiplication because it hosts massless Dirac electrons and their current flow across the Dirac node in momentum space is expected to exhibit remarkably large nonlinearity [6-13]. The nonlinearity in the current arises from drastic change of electron velocity across the Dirac node under intraband acceleration (See Supplemental Material [14]). THz HHG in graphene has, however, long evaded experimental detection. A signature was first reported in a sample consisting of 45-layer graphene at cryogenic temperature [15]. The difficulty of observing THz HHG in graphene has been attributed to fast carrier scattering, which could hamper coherent carrier transport at THz timescales [16], and thus, attention has been shifted to HHG by mid-IR excitation at higher light frequency that exceeds the scattering rate [17-20]. Followed by observation of THz THG in a 2D Dirac system at the surface of a topological insulator [21], very recently the THz HHG in graphene was clearly demonstrated by using very intense light sources based on large accelerators [22]. The nonlinear coefficients in graphene was found to be much larger than typical materials [22], whereas the conversion efficiency is limited by its monolayer nature. Therefore, realizing higher conversion efficiency in a bulk material is desired for practical application. Another issue to be resolved is the origin of THz HHG in graphene, which was explained by a thermodynamic model where electrons are assumed to be in quasi-equilibrium with repeating heating/cooling processes very rapidly within THz timescale [22,23]. Such a picture of incoherent electron dynamics is an essentially different mechanism from that originally proposed as a source of large nonlinearity in massless Dirac systems [6-13]. A recent time-resolved ARPES for a topological insulator has reported ballistic current of Dirac electrons driven by THz pulse [24]. Getting deeper insight into the microscopic dynamics of Dirac electrons in the high field regime under the influence of scattering is essential as well as is finding other bulk candidates for efficient THz HHG.

In this Letter, we investigate THz nonlinear transmission in thin films of the three-dimensional (3D) Dirac semimetal $Cd_3As_2$ and report the observation of THz harmonic generation to the extent that it is observable with a tabletop light source at room temperature. We also perform time-resolved THz spectroscopy to reveal the ultrafast dynamics of Dirac fermions during the THG process. These results provide evidence for the remarkable THz nonlinearity originating from coherent intraband acceleration of Dirac electrons, as shown in Fig. 1(a), which is in good agreement with the theoretical calculation including the scattering. Our results provide deeper understanding of Dirac electron current and show the path forward to design efficient frequency-conversion devices that are based on the remarkable features of massless Dirac electrons.

Samples consisting of (112) $Cd_3As_2$ thin films with 240-nm thickness are epitaxially grown on a GaSb buffer layer on a GaAs substrate [25,26]. The band structure of $Cd_3As_2$ has two Dirac nodes near the $\Gamma$ point and quasi-linear band dispersion up to the energy scale of ~1 eV [27,28]. Considering the low concentration and the small effective mass of carriers, the Fermi energy $E_F$ is estimated to be as small as ~50 meV above the Dirac nodes [29]. We performed THz time-domain spectroscopy to evaluate the linear response [14]. Figures 1(b) shows the real and imaginary THz conductivity for the film with the fitted results using the Drude model. We found that the scattering time $\tau_s$ is as long as 145 fs at 300 K, which is longer than usual metals and graphene. Such coherent transport is often observed in 3D Dirac or Weyl semimetals with prohibited backward scattering [30]. Interestingly, $\tau_s$ is longer at higher temperature [14], which might be accounted for by the screening of scattering by long-range Coulomb disorders [31-33].

For nonlinear spectroscopy intense THz pulses are generated by the tilted-pulse-front scheme [34,35] with bandpass filters at $f=\omega/2\pi$=0.8 THz [14]. The incident peak field is 31 kV/cm in atmosphere. Figure 1(c) shows power spectra of the THz pulse after transmitting through the $Cd_3As_2$ sample. A sharp peak of THG is clearly observed at $3f$=2.4 THz. Clearer data of the fifth harmonics are also shown in Supplemental Material [14]. For comparison, we examined monolayer graphene on a SiC substrate with the same excitation condition and also found the THG signal there as well. In contrast, a reference substrate of GaSb/GaAs shows no THG. Figure 1(d) shows the THG field amplitude as a function of the incident fundamental field strength $E_{pump}$. The data scale as $\sim E_{pump}^3$ for weaker field regimes, but they saturate for stronger excitation, showing nonperturbative behavior. In the perturbative regime for graphene, we estimated the nonlinear coefficient $\chi^{(3)}$ as ~$10^{-9}$ $m^2V^{-2}$. These results for graphene are consistent with the previous literature [22] reporting $\chi^{(n)}$ to be 7-18 orders of magnitude larger than typical materials. Compared to graphene, the transmitted fundamental wave for $Cd_3As_2$ is smaller because of the Fresnel loss; the most part of the incident THz wave is reflected at the metallic surface and only the field of 6.5 kV/cm is applied inside the $Cd_3As_2$ film. Nevertheless, the detected THG for $Cd_3As_2$ is stronger than that for graphene, which exhibits much greater conversion efficiency in $Cd_3As_2$. The THG field amplitude is as strong as 100 V/cm. The Fresnel reflection loss can be avoided by anti-reflection coating on the film or directly applying the field via contact electrodes, which will generate further stronger harmonics as a realistic frequency conversion for ultrafast electronic application.

To explore the microscopic picture of the efficient HHG mechanism in $Cd_3As_2$, we performed two types of THz pump-THz probe spectroscopy (TPTP) [5,36,37]. See details in Supplemental Material [14]. Figure 2(a) shows a schematic of TPTP with a broadband short pump pulse with a maximum peak field of 50 kV/cm in air and center frequency of 2 THz. Figure 2(b) shows the probe THz waveform $E_{probe}$ as a function of the probe delay $t_2$ against the gate pulse for the electro-optic sampling. Figure 2(c) shows a 2D plot of the change of the probe waveform by THz pump, $\delta E_{probe}$. With fixing $t_2$ at the peak of the probe pulse, we plot $\delta E_{probe}$ as a function of the pump delay $t_1$ in Fig. 2(d), which provides information of ultrafast dynamics induced by the THz pump. $\delta E_{probe}$ grows just after the pump arrives and then decays slowly. The small dip at $t_1 \sim 8$ ps originates from the reflected THz pump pulse from the back surface of the substrate. Here the probe polarization is perpendicular to the pump ($\boldsymbol{E}_{probe} \perp \boldsymbol{E}_{pump}$), but we confirmed that the result is similar with the case of $\boldsymbol{E}_{probe} \parallel \boldsymbol{E}_{pump}$ [14]. By Fourier transformation, the time evolution of the conductivity spectra $\sigma(\omega, t_1)$. Figures 2(e) and 2(f) show the real and imaginary parts at $t_1 = 5$ ps. Except for the very fast timescale within the pump pulse duration ($t_1 < 2$ ps), the conductivity spectra are reasonably fitted by the Drude model, and the fitting parameters of the carrier density $N$ and the scattering time $\tau_s$ are plotted in Figs. 2(g) and 2(h) as a function of $t_1$. The increase of $N$ implies that electrons in the valence band are excited to the conduction band by the THz pump. According to the increase of the carrier density and electron temperature, $\tau_s$ also increases and then slowly decays. It could be accounted for by the screening effect of long-range Coulomb impurity scattering [31-33] and is consistent with the temperature dependence of the equilibrium results [14]. By fitting the relaxation of carrier density with an exponential decay considering multiple reflections of the pump pulses, the relaxation time $\tau_R$ was evaluated to be ~8 ps, which is consistent with previous pump-probe studies for $Cd_3As_2$ with near- and mid-IR excitation [38-40]. In contrast to the case of rapid cooling in graphene (< 1 ps) [41], the relaxation time $\tau_R \sim 8$ ps in $Cd_3As_2$ is much longer than our pump THz field cycle, indicating that, unlike graphene, the rapid heating/cooling cycles do not occur during the THz wave irradiation. In fact, fort this slow relaxation, we performed the thermodynamic-model calculation in Supplemental Material [14] and confirmed that this model cannot reproduce the efficient THG with realistic parameters.

The nonthermal mechanisms of HHG in solids have been discussed in recent literatures in terms of the interband polarization and intraband current [4]. The contribution of interband polarization on HHG, which might be dominant in semiconductors with mid-IR excitation, cannot dominate for THz frequencies since the dephasing time is much shorter than the field cycle and therefore the short-lived polarization cannot emit low-frequency light [42,43]. Note that the interband transition could occur in the Dirac semimetal with supplying additional population of carriers to contribute nonlinear current. However, the increase of the carrier density observed in Fig. 2(g) was only 20 % even for the broadband intense pump pulse with the center frequency of 2 THz and the peak field of 50 kV/cm. Since $E_F$~50 meV is larger than the incident THz photon energy (~several meV), the interband transition is not efficient due to the Pauli blocking. For the THG experiment in Fig. 1(c), the interband excitation would be further small because of the lower frequency of 0.8 THz and weaker field strength of 31 kV/cm. Therefore, the results indicate that the interband transition is not a major effect for efficient THG and that the intraband current is the main source of this nonlinearity. A similar situation in a doped Dirac system has been discussed in a previous calculation [12].

To obtain further insight into the *dynamical evolution during the THz THG process*, we also performed another TPTP with quasi-monochromatic pump waveform as schematically shown in Fig. 3(a) [14]. Figure 3(b) shows the pump waveform with $f$=0.8 THz and the change of the probe field $\delta E_{probe}$ as a function of $t_1$ during the quasi-monochromatic pump irradiation. For the case of $\boldsymbol{E}_{probe} \parallel \boldsymbol{E}_{pump}$, an oscillation signal is discerned on the slow-rising incoherent carrier excitation. Figure 3(c) shows its Fourier component with a peak frequency at 1.6 THz. This result means that the THz conductivity oscillates in time with the frequency $2f$ during the pump irradiation. Such a coherent oscillation under a phase-stable pump field can be observed generally in pump-probe experiments with subcycle time resolution [44]. Note that the "$2f$-oscillation" in the pump-probe signals directly corresponds to the THG in the transmission of the pump because the nonlinear $3f$ current arises from the oscillating conductivity with frequency $2f$, as also studied in superconductors [5] (See more details in Supplemental Material [14]). Importantly, for the case of $\boldsymbol{E}_{probe} \perp \boldsymbol{E}_{pump}$, the $2f$-oscillation signal is hardly seen in time domain and quite small compared to the case of $\boldsymbol{E}_{probe} \parallel \boldsymbol{E}_{pump}$. This is in a stark contrast to the case of superconductors [5], where the $2f$-oscillation was clearly observed in $\boldsymbol{E}_{probe} \perp \boldsymbol{E}_{pump}$; there it is because the *s*-wave order parameter oscillates in time, which results in the isotropic change of the conductivity [45]. If the THG is described by the thermodynamic model [22], the change of the conductivity is also isotropic. Our results for $Cd_3As_2$ are, however, clearly distinct from these models and shows that the nonlinearity manifests itself almost only for the pump polarization direction. The result can be well explained by the intraband acceleration of electrons. When electrons are accelerated by the pump field, the electron distribution function moves rapidly back and forth in the momentum space along with the pump polarization direction, but it appears as if "nothing happens" in other probe polarization directions, which is observed as the anisotropic $2f$-oscillation.

Both of our TPTP experiments strongly indicate that the electrons are quite nonthermal under THz field and therefore the intraband current of accelerated electrons is the main source of the extremely efficient THG. Here we calculated the nonlinear current considering (i) the intraband acceleration in a simple linear dispersion model with $\tau_s$=145 fs and (ii) the thermodynamic model with $\tau_R$=8 ps [14]. Figure 4(a) shows the HHG spectra for both models with the incident field $E_{pump}$=31 kV/cm. Figure 4(b) shows the calculated THG amplitudes for both models as a function of $E_{pump}$ in compassion with the experimental data. The coherent acceleration model can reproduce the value of the THG amplitude within the factor of 2 and its saturating behavior even in the simplified model. On the other hand, the thermodynamic model strongly saturates with smaller THG amplitude because the heating piles up due to the long relaxation time, which cannot account for the experimental results.

One might think that such a nonlinear current from coherently accelerated electrons would be significantly suppressed by the scattering since the scattering rate $1/\tau_s$~7 THz is much faster than the pump frequency of 0.8 THz. However, note that even in this case the nonlinear current by coherent acceleration can appear as long as *the electrons are driven into an unbalanced momentum distribution before the scattering*. Therefore, the comparison between $1/\tau_s$ and $f$ is not important, and we should compare $\tau_s$ with the time for building up the unbalanced electron momentum distribution. As an extreme example, when ac electric field $E_0 \cos(2\pi f t)$ is applied inside the film, the time required for accelerating the electrons from the Fermi surface to the Dirac node can be estimated by the acceleration theorem [14] as

$$t' = \frac{1}{2\pi f} \sin^{-1}\left(\frac{2\pi f}{E_0} \frac{E_F}{ev_F}\right),$$

where $v_F$=0.93×10$^6$ m/s is the Fermi velocity [27]. It gives $t'$~85 fs for $E_0$=6 kV/cm and $E_F$=50 meV, which is shorter than $\tau_s$=145 fs. Importantly, $t'$ is not sensitive to the frequency $f$. Therefore, the electrons are coherently accelerated by several-kV/cm THz field to form a quite nonthermal momentum distribution before the scattering and contribute to the nonperturbative HHG even for the low-frequency driving force.

Figure 4(c) shows the calculated result of the THG amplitude as a function of $\tau_s$ for the peak fields of 3 and 6 kV/cm inside the film and $f$=0.8 THz. If $\tau_s$ is longer than ~85 fs, the THG amplitude strongly saturates into the nonperturbative regime and behaves as if in the long-$\tau_s$ limit. Therefore, $\tau_s$~150 fs in this material is "long enough" for intraband acceleration driven by a few-kV/cm field at 0.8 THz. Figure 4(d) shows the 2D plot of electron distribution in the momentum space driven by the field in comparison with equilibrium. A few-kV/cm THz field inside the film induces substantial unbalance of the intraband electron distribution in the pump polarization direction. Note that such a nonthermal population of Dirac electrons has been also recently demonstrated in a time-resolved ARPES in a topological insulator driven by a THz field [24].

In conclusion, we observed the extremely efficient THz THG in the Dirac semimetal $Cd_3As_2$. Our TPTP experiments as well as the calculations based on the coherent and incoherent models clearly demonstrate that the THz field coherently accelerates Dirac electrons. In spite of the presence of the scattering, several-kV/cm THz field can drive the system into a nonthermal, unbalanced momentum distribution before the scattering, which generates nonperturbative nonlinear current. Even longer scattering times might be achieved in other topological semimetals with massless dispersion, which will show even stronger frequency conversion perhaps in sub-kV/cm field strength. The results in this work opens a new pathway to develop a novel frequency convertor in THz frequency based on Dirac semimetals.

**Acknowledgments**

The experiment was supported by JST PRESTO (Grant No. JPMJPR16PA) and in part by JSPS KAKENHI (Grants Nos. JP19H01817, JP19K15462, and JP18K13495). R.M. also acknowledges support by Attosecond lasers for next frontiers in science and technology (ATTO) in Quantum Leap Flagship Program (MEXT Q-LEAP). B.C. and N.P.A. were supported by NSF EFRI 2-DARE 1542798 and NSF DMR 1905519. N.P.A. acknowledges additional support from the JSPS International Research Fellows Program. T.S. and S.S. acknowledge support through the Vannevar Bush Faculty Fellowship program by the U.S. Department of Defense (Grant No. N00014-16-1-2814).

B. C. and N. K contributed equally to this work.

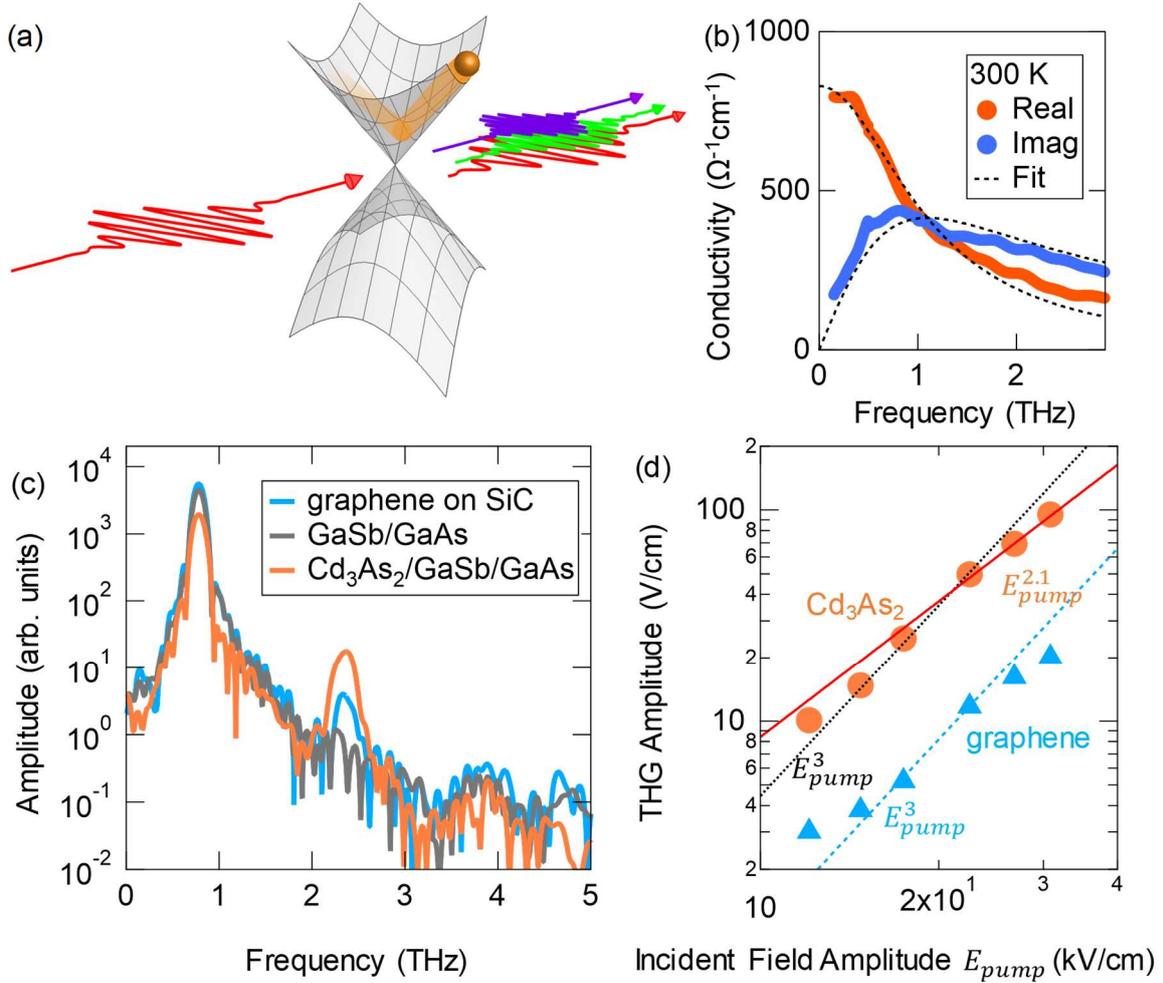

**Figure 1.** (a) Schematic of the HHG in Dirac electron system. (b) Real- and imaginary-part THz conductivity spectra at 300 K. The dotted curves are the results of fitting. (c) Amplitude spectra of transmitted pump THz pulses for the $Cd_3As_2$ film, a reference substrate, and graphene on SiC with $E_{pump}$=31 kV/cm. (d) THG field amplitude as a function of $E_{pump}$. The solid line for the $Cd_3As_2$ data is a fitted result with function $\propto E_{pump}^\alpha$ with $\alpha$=2.1 for stronger field. The dashed lines are fitted with $\propto E_{pump}^3$ to show the perturbative regimes.

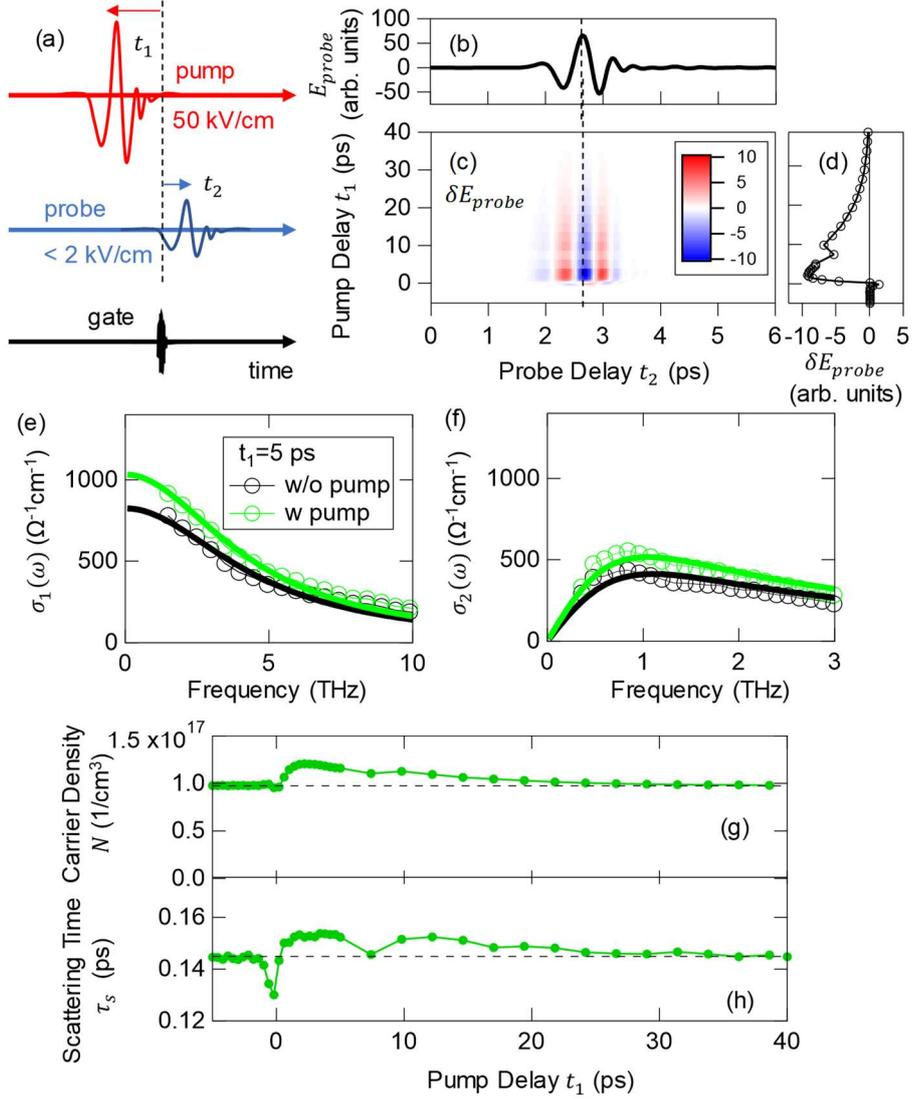

**Figure 2.** (a) Schematic of broadband short THz pump-THz probe spectroscopy. (b) Probe pulse waveform $E_{probe}$ as a function of $t_2$. (c) 2D plot of the change of the probe electric field $\delta E_{probe}$ by the THz pump as functions of $t_1$ and $t_2$. (d) $\delta E_{probe}$ with a fixed delay of $t_2$ at the peak of the probe indicated by the broken line in (c). (e)(f) Open circles show the real and imaginary parts of the transient conductivity spectra at $t_1=5$ ps in comparison with data in equilibrium. The solid curves show the fitting with the Drude model. (g)(h) The carrier density $N$ and scattering time $\tau$ obtained by the fitting in (e) and (f) as a function of the pump delay $t_1$.

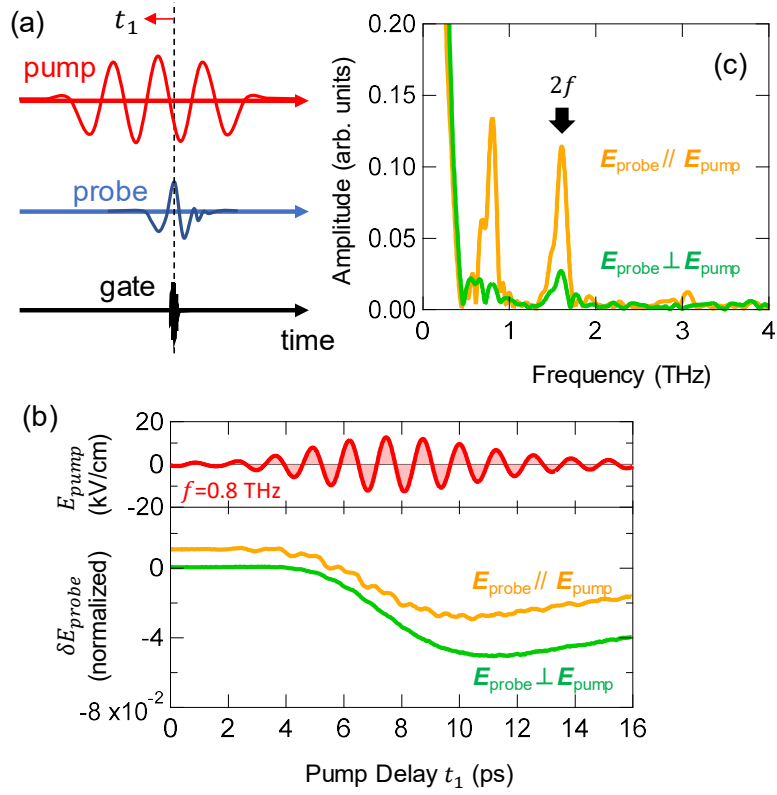

**Figure 3.** (a) Schematic of quasi-monochromatic THz pump-THz probe measurement. (b) The upper figure shows the pump THz waveform with $f$=0.8 THz. The lower figure shows $\delta E_{probe}$ as a function of $t_1$ for the probe polarizations parallel and perpendicular to the pump. The data are normalized at the peak value of probe field and shown with offset for clarity. (c) Fourier components of $\delta E_{probe}$ on the interband excitation signal in (b). The arrow indicates the oscillation component at $2f$=1.6 THz. The peak at $f$=0.8 THz is ascribed to an artifact [14].

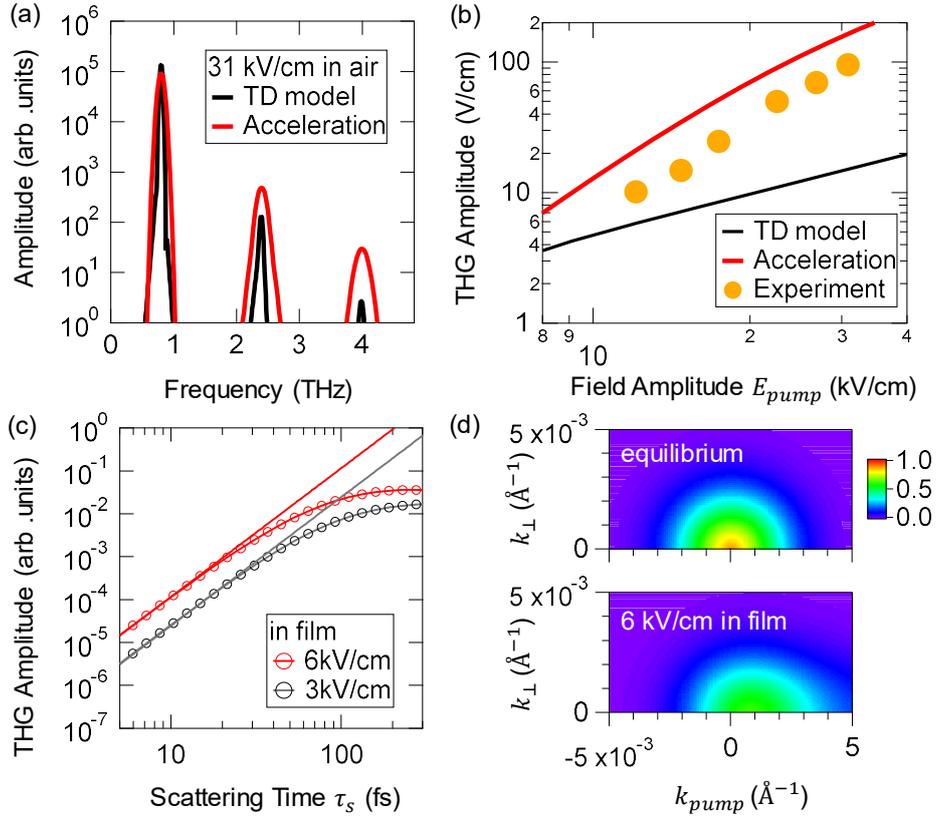

**Figure 4.** (a) Calculated amplitude spectra from the thermodynamic (TD) model and the intraband acceleration model for $f$=0.8 THz. (b) THG amplitudes as a function of the field strength. The black and red curves show the results of the TD model and the intraband acceleration model and the circles show the experimental data. (c) THG amplitude as a function of $\tau_s$ for $f$=0.8 THz and peak field of 3 and 6 kV/cm inside the film. The solid lines show power-law fittings to clearly see saturation. (d) 2D plot of the electron distribution in the momentum space parallel and perpendicular to the pump polarization, as denoted by $k_{\text{pump}}$ and $k_\perp$, respectively. The top and bottom panels correspond to equilibrium and under the field of 6 kV/cm inside the film at $f$=0.8 THz.

# Supplemental Material for
# Efficient Terahertz Harmonic Generation with Coherent Acceleration of Electrons in the Dirac Semimetal Cd$_3$As$_2$

**Methods**

    1. Sample preparation and characterization

    2. Setup for THz time-domain spectroscopy for linear response

    3. Setup for THz THG experiment

    4. Setup for broadband short THz pump-THz probe spectroscopy

    5. Setup for quasi-monochromatic THz pump-THz probe spectroscopy

**Additional data**

    6. Third and fifth harmonics for different samples and excitation frequencies

    7. $2f$-oscillation in the pump-probe signal during THG process

    8. Additional data for $2f$-oscillation in the pump-probe signal

**Calculations**

    9. Origin of nonlinearity in the coherent acceleration of Dirac electrons

    10. Calculation of nonlinear current by intraband acceleration

    11. Calculation in the thermodynamic model

## Methods

### 1. Sample preparation and characterization

    Cd$_3$As$_2$ films were grown by molecular beam epitaxy on epitaxial GaSb buffer layers on GaAs (111)B substrates with an 1° miscut towards $\langle \bar{1}\bar{1}2 \rangle_{\text{GaAs}}$. The buffer layer and Cd$_3$As$_2$ film shows streaky reflective high-energy electron diffraction patterns throughout the growth, indicating flat surface morphologies. X-ray diffraction patterns were measured in a *Panalytical MRD PRO Materials Research Diffractometer*, using Cu K $\alpha$ radiation. Thicknesses of the GaSb (150 nm) and Cd$_3$As$_2$ layers were determined from thickness fringes around the GaSb 111, and Cd$_3$As$_2$ 224, diffraction peak, respectively (see Fig. S1). Further details on the growth can be found in previous literatures [S1,S2].

    For comparison, a commercial electron-doped graphene (monolayer coverage 92%) epitaxially grown on a 4H-SiC substrate was also used (Graphene Supermarket).

### 2. Setup for THz time-domain spectroscopy for linear response

    THz conductivity spectra $\sigma(\omega) = \sigma_1(\omega) + i\sigma_2(\omega)$ in the Cd$_3$As$_2$ films were measured in a typical THz time-domain spectroscopy setup using a mode-locked Ti:Sapphire laser with 800-nm center wavelength, 100-fs pulse duration, and 76-MHz repetition rate (Tsunami,

Spectra Physics). THz pulses were generated from an interdigitated photoconductive antenna on a GaAs substrate with 50-V biased voltage and 50-kHz modulation frequency. The transmitted THz pulses were detected by the photoconductive bow-tie antenna for low-frequency (0.15-0.6 THz) and by the electro-optical (EO) sampling in a (110) GaP crystal for high-frequency (0.6-3.5 THz), which are smoothly connected. The complex transmittance of the sample in the thin-film approximation is written as

$$t(\omega) = \frac{1}{1 + n_s + Z_0 \sigma(\omega) d} \frac{4 n_s}{1 + n_s} e^{i\Phi(\omega)},$$

where $Z_0$=377 Ω is vacuum impedance, $d$ is thickness of the $Cd_3As_2$ film, $n_s$=3.3 is the complex refractivity of the GaSb/GaAs substrate, and $\Phi(\omega) = (n_s d_s - d - d_s)\omega/c$. Here $d_s$=0.4 mm is the substrate thickness. Because the refractive indices of the buffer layer GaSb and the substrate GaAs is similar, we treated them as a single substrate.

Real and imaginary parts of the conductivity spectra are simultaneously fitted with the Drude response function;

$$\sigma(\omega) = 1 - \frac{Ne^2}{m^* \epsilon_0} \frac{1}{\omega^2 + i\omega \tau_s^{-1}} - \frac{i\epsilon_0 \omega \omega_p^2}{\omega_0^2 - \omega^2 - i\omega\Gamma},$$

where $N$ is the carrier density, $\tau_s$ is the scattering rate, and $m^*$=0.03$m_e$ is the effective electron mass measured by magneto-THz spectroscopy [S3]. The Lorentzian function in the second term was included for low-temperature data below 100 K to describe the spectral feature of the phonon at $\omega_0/2\pi$=0.56 THz with the oscillator strength $\omega_p/2\pi$=4.5 THz [S3].

Figures S1(d) and S1(e) show the results of optical conductivity spectra at various temperatures with the results of fitting. Figure S1(f) shows the scattering time $\tau_s$ obtained from the fitting as a function of temperature. $\tau_s$ becomes shorter at low temperatures. This temperature dependence is counterintuitive in usual metals but might be ascribed to a peculiar scattering mechanism of Dirac electrons; if long-range Coulomb impurity potentials dominate the scattering, they are screened more effectively by electrons at higher temperature [S4-S6].

## 3. Setup for THz THG experiment

Figure S2 shows our setup for THz THG. Output of a Yb:KGW-based regenerative amplified laser system (Pharos, Light Conversion) with the center wavelength of $\lambda_0$=1028 nm, the repetition rate of 3 kHz, the pulse energy of 2 mJ, and the pulse width of 255 fs was divided (i) for generation of intense THz pump pulse and (ii) for the gate pulse to detect the THz electric field waveform in time domain. The gate pulse width was compressed to ~50 fs by spectral broadening.

The intense THz pump pulse was generated by optical rectification in a cooled-down $LiNbO_3$ crystal with tilted-pulse-front technique [S7, S8]. A remnant of the optical pulse was blocked by a black polyethylene film. We used two wire-grid polarizers; the first one for control of the field strength and the second one for setting the field polarization vertically.

Metal-mesh THz band-pass filters for $f$=0.4, 0.6, or 0.8 THz were inserted into the beam path of the pump THz pulse to make a quasi-monochromatic pump pulse with the frequency $f$. After transmitting the sample, the pump THz pulse was detected by the EO sampling in a (110) GaP crystals with 0.3-mm or 2-mm thickness. The inset in Fig. S2 shows the calculated detection efficiency considering the phase-matching condition and dispersion in GaP with the 1028-nm gate pulse. The THz pulse can be detected efficiently up to ~3.5 THz for the 2-mm GaP crystal. Before the EO sampling, Si wafers were inserted as attenuators to keep $\Delta I/I$ less than 0.02 for linearity of the balanced detection signal. By scanning the delay of the pump $t_1$, the time-domain data were obtained and Fourier-transformed with a window function.

The power spectra of the pump pulse with the field strength of 31 kV/cm were shown in the Figure S3(a)-S3(c) with logarithm scale, linear scale, and extended linear scale. The spectra have a single sharp peak with the well-defined frequency of 0.78 THz, which we describe as $f$=0.8 THz in this paper unless otherwise mentioned. The full width of half maximum is $\Delta f$=0.78 THz and $\Delta f/f$ is as small as 0.1.

The pump THz field strength in air was estimated at the position of the sample using the EO sampling in a (110) GaP crystal with the thickness of $L$=0.3 mm as

$$E_{pump} = \frac{n_{THz}+1}{2} \frac{\lambda_0}{2\pi n_0^3 r_{41} L} \frac{\Delta I}{I},$$

where $n_0$=3.11 and $n_{THz}$=3.34 is the refractive index of GaP at $\lambda_0$ and at THz frequency, respectively, $r_{41}$=0.97 pm/V is the EO coefficient $\Delta I/I$ is the balanced detection signal. The field strength in this paper is defined as the value in air unless otherwise mentioned. The field strength inside the thin film is described as

$$E_{pump}^{InFilm} = \frac{1}{1 - \frac{n-1}{n+1}\frac{n-n_s}{n+n_s}e^{i2\frac{n\omega}{c}d}} \frac{2}{n+1}\left(1 + \frac{n-n_s}{n+n_s}e^{i2\frac{n\omega}{c}d}\right) E_{pump},$$

where $n$ is the refractive index of the Cd$_3$As$_2$ film evaluated by our THz time-domain spectroscopy. The factor $|E_{pump}^{InFilm}|/|E_{pump}|$ is 0.19 and 0.21 for $f=\omega/2\pi$=0.6 and 0.8 THz, respectively, for the 240-nm film for linear response regime.

Note that this factor could alter in the nonlinear response regime. Figure S3(d) shows the experimentally-observed transmittance of the fundamental component as a function of the incident field strength for the Cd$_3$As$_2$ film and graphene. The transmittance for the Cd$_3$As$_2$ film sample decreases slightly for strong field. This nonlinearity is ascribed to the excitation of carriers by the intense THz pulse, which can increase absorption and reflection of the THz pulse itself. The decrease of the transmittance is only 10% even at $E_{pump}$=31 kV/cm, which indicates that the effect on the field strength inside the film is also less than 10%. In this work we ignored this nonlinearity and evaluated the field strength inside the film as $E_{pump}^{InFilm}$=6.5 kV/cm for $E_{pump}$=31 kV/cm as it is in linear response regime.

Figures S3(e) and S3(f) show the time-domain waveforms and power spectra of the quasi-monochromatic pump pulse for various field strengths controlled by the wire-grid polarizer. Figures S3(g) and S3(h) show the same data normalized for comparison. Spectral change is negligible for all the field strength. In time domain, the waveforms are very same from 31 kV/cm to 7.1 kV/cm. For much weaker fields of 1.8 or 0.6 kV/cm, small shifts of the carrier-envelope phase (CEP) can be discerned. Such a CEP shift is known to appear when two wire-grid polarizers are placed in close to the crossed-Nicol configuration. Since these field strengths are small enough and the waveform is multicycle, the interaction is insensitive to the carrier-envelop phase and therefore the shift of CEP is not relevant to the discussion in this work.

## 4. Setup for broadband short THz pump-THz probe spectroscopy

To investigate *the dynamics of carrier relaxation* induced by the intense THz pump, we performed THz pump-THz probe spectroscopy. Figure S4 shows our setup and it is basically the same with Fig. S2 unless otherwise mentioned. Here we removed the THz band-pass filters to make a broadband short THz pump pulse with the pulse duration <2 ps for increasing time resolution and pump pulse intensity. The pump waveform and power spectrum are shown in Figs. S5(a) and S5(b), respectively.

The compressed 50-fs pulse was further split for the gate pulse and for the probe THz pulse generation. The probe THz pulse was emitted from a (110) GaP crystal with 0.5-mm thickness. To combine the beam path of the THz pump and THz probe, we used a home-made aluminum cylindrical block with 2-inch outer and 1-inch inner diameters as shown in the inset of Fig. S4. It was cut at the angle of 45 degrees with a mirror-polished surface. The pump THz pulse transmits the 1-inch hole with a reduced field strength by a factor of ~2, while the probe THz pulse is reflected at the donut-shaped mirror surface, then both pulses collinearly irradiate on the sample.

We carefully checked spatial overlap of the pump and probe pulse. Since the numerical aperture of the probe is bigger than the pump and the donut-shape beam is more tightly focused than a gaussian beam, the probe pulse spot size can be smaller than the pumped area. We performed the knife-edge method and evaluated the radius and center position of the pump and probe pulses as shown in Figs. S5(c) and S5(d), respectively. The probe pulse was well overlapped with smaller spot size for all the frequency window, ensuring proper alignment for pump-probe spectroscopy.

After transmitting through the sample, we removed the Si attenuators and placed a metal plate with 1-inch diameter to block only the pump with smaller diameter. The polarization direction of the probe THz pulse is freely selectable using a wire-grid polarizer in the probe beam path and rotation of the GaP crystal. Optical choppers were used for 1500- and 750-Hz modulations of the pump and probe pulse, respectively. The delay times of the pump and probe, $t_1$ and $t_2$, respectively, were controlled independently. The data were taken for each shot of the pulsed laser one by one and analyzed in a similar way with a previous literature [S9], and then we obtained the probe field waveform with and without the pump, $E'_{probe}(t_1,t_2)$ and $E_{probe}(t_1,t_2)$, respectively.

The change of the probe field $\delta E_{probe}(t_1,t_2) \equiv E'_{probe}(t_1,t_2) - E_{probe}(t_1,t_2)$ is plotted in Fig. 2(c) in the two-dimensional time domain. For each pump delay $t_1$, the complex optical conductivity with the pump $\sigma'(\omega)$ shown in Figs. 3(e) and 3(f) were calculated from the following relation:

$$\frac{E_{probe}(\omega)}{E'_{probe}(\omega)} = \frac{1 + n_s + Z_0\sigma'(\omega)d}{1 + n_s + Z_0\sigma(\omega)d}.$$

Compared to previous THz pump-THz probe measurement [S9-S11], this setup using the donut-shaped mirror has an advantage of selectable polarization direction for the probe THz pulse. Figures S5(e) and S5(f) show the results of the pump-probe signal $\delta E_{probe}(t_1)$ for 280-nm film at room temperature with the probe polarization parallel and perpendicular to the pump, respectively. The probe delay $t_2$ was fixed at the peak of $E_{probe}(t_2)$ and the data were normalized with the value of the peak. Figures S5(e) and S5(f) give quantitatively similar results, which indicates that the carrier excitation and relaxation dynamics can be observed in both polarization directions.

### 5. Setup for quasi-monochromatic THz pump-THz probe spectroscopy

To investigate *the dynamics during the THG process*, we performed the THz pump-THz probe measurement with quasi-monochromatic THz pump. The setup is very similar to the one in previous section except for using the band-pass filters.

Compared to the previous section, more careful data taking was needed since this measurement was performed to investigate the dynamics when the pump and probe THz pulses overlap in time. Even though the pump THz pulse was blocked by the metal plate, the remnant of the pump pulse was also detected in the EO sampling of the probe pulse, which could cause artifacts originating a nonlinear effect in the EO sampling.

Before the measurement, we checked the pump-probe signal $\delta E_{probe}$ without the sample, and then found that an oscillating signal with $f=0.8$ THz appeared in $\delta E_{probe}$ which can be ascribed to the artifact of the THz pump at the EO sampling in the GaP crystal. This artifact appeared especially for $\boldsymbol{E}_{probe} \parallel \boldsymbol{E}_{pump}$, and we could not completely eliminate it. Then we put the sample and performed the THz pump-THz probe spectroscopy. When the sample is used, the $2f$-oscillation appeared only in the configuration of $\boldsymbol{E}_{probe} \parallel \boldsymbol{E}_{pump}$ as shown in Fig. 3(c), indicating that it undoubtedly originates from the nonlinear interaction in the sample.

**Additional data**

## 6. Third and fifth harmonics for different samples and excitation frequencies

We used samples with different thicknesses of 140, 240, 280, and 340 nm for THG experiment for the excitation at $f=0.8$ THz at room temperature and found that all the samples show strong THG. Figure S6(a) shows the angle dependence of the THG intensity measured by rotating the 280-nm sample. The polarization dependence of the THG intensity was almost isotropic. Figure S6(b) shows the Fourier spectra of the transmitted pulses for the 280-nm sample with different excitation frequencies of $f=0,4$, 0.6, and 0.8 THz, and we confirmed that the THG signal is observed for every condition.

For excitation at 0.6 THz, we inserted additional band-pass filters for 3.0 THz after transmitting through the 240-nm sample to block the fundamental wave. The peak field was 13.8 kV/cm in air, which was reduced to 2.62 kV/cm in the thin film. Figure S6(c) shows the spectra of the transmitted pulses, and the signals of fifth harmonic generation (FHG) were observed at around $5f \sim 3.0$ THz. Figure S6(d) shows the FHG intensity as a function of fundamental wave intensity. The signal shows clear saturation in the nonperturbative regime.

## 7. $2f$-oscillation in the pump-probe signal during THG process

When an intense monochromatic pump electric field $E_{pump}(t_1) = E_0 e^{i\omega t_1}$ is applied in a sample with inversion symmetry, the current density including perturbative nonlinear response is described as

$$j_{pump}(t_1) = \int_{-\infty}^{t_1} dt' [\sigma(t_1 - t')E_{pump}(t') + \sigma^{(3)}(t_1 - t')E_{pump}(t')^3 + \cdots]. \quad (1)$$

Here we neglect the interband transition for simplicity. The discussion below is independent of gauge. Transmitted electric field is proportional to time derivative of the current density $j_{pump}(t_1)$. In the frequency domain, Eq. (1) up to the third order becomes

$$j_{pump}(\omega) = \sigma(\omega)E_{pump}(\omega) + \sigma^{(3)}(3\omega; \omega + \omega + \omega)E_{pump}(\omega)^3. \quad (2)$$

The nonlinear conductivity $\sigma^{(3)}$ in the second term of the right side is the source of THG observed in the transmittance of the pump in Fig. 1(c).

In the pump-probe spectroscopy in Fig. 3, another probe electric field $E_{probe}(t_2)$ is also applied in the sample during the monochromatic pump pulse irradiation. Delays of the pump and probe pulses, $t_1$ and $t_2$, respectively, are controllable independently. The electric field of transmitted probe pulse is emitted from the current density

$$\begin{aligned} j_{probe}^{pump}(t_1, t_2) &= \sigma E_{probe}(t_2) + \sigma^{(3)} E_{pump}^2(t_1) E_{probe}(t_2) \\ &= (\sigma + \sigma^{(3)} E_0^2 e^{i2\omega t_1}) E_{probe}(t_2). \end{aligned}$$

Here for simplicity we assumed an instantaneous response, *i.e.*, we regarded $\sigma$ and $\sigma^{(3)}$ as time-independent constants. We measured the change of the probe electric field $\delta E_{probe}(t_1, t_2)$, which is related to the difference of the current density $\delta j(t_1, t_2) \equiv j_{probe}^{pump}(t_1, t_2) - j_{probe}(t_2) = \sigma^{(3)} E_0^2 e^{i2\omega t_1} E_{probe}(t_2)$. By fixing the delay $t_2$ at the peak of the probe waveform and by scanning the pump delay $t_1$, the pump-probe signal $\delta E_{probe}(t_1)$ can be measured, which is proportional to $\sigma^{(3)}$ and oscillates in time with twice the incident frequency. Importantly, both of the THG and the 2f-oscillation originate from the third-order nonlinear conductivity $\sigma^{(3)}$. In other word, the very same phenomenon induced by the pump is observed from different aspects in the transmission of the pump and in the pump-probe signal.

Note that in a precise form, $\sigma$ and $\sigma^{(3)}$ are the response functions depending on time and therefore there are retardation and finite time resolution between the field and the current. But the pump-probe measurement still works well to resolve the oscillation in THz timescale as discussed in the Supplemental Material in a previous literature [S10].

More generally, the third-order response function $\sigma^{(3)}$ is a fourth-rank tensor depending on polarizations such as $j_l^{(3)} = \sigma_{ijkl}^{(3)} E_i E_j E_k$ $(i,j,k,l = x,y,z)$. While the THG experiment detects $\sigma_{iiii}^{(3)}$ for the pump polarization along $i$, the pump-probe measurement can also investigate $\sigma_{iikk}^{(3)}$ $(k \neq i)$ when the probe polarization $k$ is not parallel to the pump. In general, $\sigma_{xxxx}^{(3)}$ and $\sigma_{xxyy}^{(3)}$ can be different since they indicate that how the nonlinearity induced by the $i$-polarized pump appears in $i$ and $k$ direction, respectively $(k \neq i)$. Therefore, the "$2f$-oscillation" measurement provides more detailed information for ultrafast dynamic in materials right in the middle of nonlinear interaction. In our pump-probe measurement in Figs. 3(b) and 3(c), the "$2f$-oscillations" were clearly observed for $k = i$ but very small or not observed for $k \neq i$. It means that, during the THG process, the electron system is drastically changing along the pump direction $i$ but nothing happens along the direction $k$ ($\neq i$) except for the incoherent carrier excitation. It is in a stark contrast to the previous experiment in a superconductor[5] or the thermodynamic model proposed in graphene[21], where the change of the conductivity should appear isotropically in other polarization directions. In contrast, in the picture of coherent acceleration in momentum space, the drastic change of the electron distribution function occurs only along with the pump polarization direction, as shown in the simulation in Fig. 4(d). Therefore, the large $\sigma_{iiii}^{(3)}$ observed in THG and the absence of $\sigma_{iikk}^{(3)}$ ($k \neq i$) in TPTP revealed in this study strongly indicates that the microscopic origin of THz THG is the intraband acceleration of electrons in the $i$ direction without scattering into other momentum space.

Not only the $2f$ oscillation, another peak at the frequency $f$ was also observed in Fig. 3(c). Note that the $f$ peak does not originate from a second-order nonlinearity $\sigma^{(2)} E_0 e^{i\omega_1} E_{probe}$ since even-order harmonics were not observed in the transmission measurement because of the presence of inversion symmetry. We assigned the peak to the artifact which could occur due to nonlinearity in the EO sampling as mentioned above. We performed the same measurement without $Cd_3As_2$ and also found the appearance of the peak at $f$. But the $2f$ oscillation was observed only when $Cd_3As_2$ was used with $\boldsymbol{E}_{probe} \parallel \boldsymbol{E}_{pump}$.

## 8. Additional data for $2f$-oscillation in the pump-probe signal

Here we examined the observation of the $2f$-oscillation in the pump-probe signal more carefully. Figure S7 shows the result of TPTP with quasi-monochromatic pump pulse with center frequency $f$=0.8 THz. The left and right columns correspond to the results in cases of $\boldsymbol{E}_{probe} \parallel \boldsymbol{E}_{pump}$ and $\boldsymbol{E}_{probe} \perp \boldsymbol{E}_{pump}$, respectively.

Figure S7(a) shows the probe THz waveform. We fixed the probe delay at three different points (positive peak, zero cross, and negative peak) as indicated by broken lines. At all these points, we measured the pump-probe signal $\delta E_{probe}$ as a function of the pump delay $t_1$ as shown in Fig. S7(b). Figure S7(c) shows its Fourier components. The $2f$-oscillation of 1.6-THz frequency was clearly observed at any probe delay point in the case of $\boldsymbol{E}_{probe} \parallel \boldsymbol{E}_{pump}$. But such an oscillation was never identified in the case of $\boldsymbol{E}_{probe} \perp \boldsymbol{E}_{pump}$. We also performed the experiment with the pump frequency $f$=0.6 THz and confirmed that the results are very similar.

# Calculations

## 9. Origin of nonlinearity in the coherent acceleration of Dirac electrons

Here we briefly explain an intuitive picture of the nonlinearity in the coherent acceleration of Dirac electrons proposed in Ref. [S12]. Let us consider the case applying an ac electric field $E_x(t) = E_0 \cos(2\pi f t)$ in a crystal. Here we neglect the interband polarization and only consider the intraband process. From the acceleration theorem, the electron momentum is changed as

$$\frac{dk_x(t)}{dt} = -\frac{eE_0}{\hbar}\cos(2\pi f t). \qquad (3)$$

When the electron has a dispersion relation $\epsilon(\boldsymbol{k})$, the electron velocity is described as

$$v_x(t) = \frac{1}{\hbar}\frac{\partial \epsilon(\boldsymbol{k})}{\partial k_x}.$$

If the dispersion is parabolic, $\epsilon(\boldsymbol{k}) = a|\boldsymbol{k}|^2$, we obtain the velocity $v_x(t) \propto \cos(2\pi f t)$. Since the current is proportional to the velocity, the current has also the same frequency $f$, *i.e.*, the response is totally linear. However, in the case of massless dispersion, $\epsilon_{\boldsymbol{k}} = \pm \hbar v_F |\boldsymbol{k}|$, the velocity is obtained as $v_x(t) = -v_F \text{sgn}(\sin(2\pi f t))$, indicating that the velocity drastically changes its sign. Therefore, a square-wave-like current is induced when a cosine-type applied field, as discussed in Ref. [6]. This is the nonlinearity and manifestation of the nature of Dirac electrons.

Note that the model explained here is the simplest case. In the real material used in our work, the electrons are doped up to the Fermi energy of 50 meV. The momentum scattering can be also involved. Therefore, compared to the ideal picture above, the nonlinearity observed will be more obscured. Nevertheless, the remarkable nonlinearity was confirmed in our experiment. More detailed calculations are provided below.

One might think that such a picture of coherent intraband acceleration would be invalid in the presence of fast scattering since the scattering rate $1/\tau_s \sim 7$ THz is faster than the pump frequency $f=0.8$ THz in the present experiment. However, note that even in such a situation, the nonlinear current by coherent acceleration can appear as long as the intraband electron distribution becomes unbalanced in the momentum space before the scattering. Therefore, the important factor for the appearance of nonlinear current by coherent acceleration is, not the ratio between $1/\tau_s$ and $f$, but the ratio between $\tau_s$ and the time for building up unbalanced momentum distribution. As an extreme example, by integrating the acceleration theorem in Eq. (3), we can calculate "the time required for the electrons accelerated from the Fermi surface reaching to the Dirac node", which can be expressed as

$$t' = \frac{1}{2\pi f} \sin^{-1}\left(\frac{2\pi f}{E_0} \frac{E_F}{e v_F}\right).$$

For $E_0 = E_{pump}^{InFilm} = 6$ kV/cm and $E_F=50$ meV, we obtain $t'=85$ fs, which is shorter than $\tau_s=145$ fs. Importantly, $t'$ is not sensitive to the pump frequency $f$ and mainly determined by $E_0$. Therefore, the field strength of a few kV/cm inside the film is strong enough to coherently accelerate the electrons to make a significantly-unbalanced electron distribution in the momentum space, which causes nonperturbative nonlinear current.

**10. Calculation of nonlinear current by intraband acceleration**

Following a previous literature [S13], we have invoked the time evolution equation for the electron-density $\rho(\mathbf{k}, t)$ at time $t$:

$$\frac{\partial \rho(\mathbf{k},t)}{\partial t} = -\frac{e}{\hbar} \mathbf{E}(t) \cdot \nabla_{\mathbf{k}} \rho(\mathbf{k},t) - \frac{\rho(\mathbf{k},t) - f_{FD}(\mathbf{k})}{\tau_s}, \quad (4)$$

where $e$ ($<0$) is the electron charge, $\hbar$ the Planck constant, and $\tau_s$ the scattering time. Here we used the length gauge. We let $f_{FD}(\mathbf{k}) = \left[\exp\left(\frac{\epsilon_{\mathbf{k}} - E_F}{k_B T}\right) + 1\right]^{-1}$ denote the Fermi-Dirac distribution, and have approximated the quasi-linear band dispersion of Cd$_3$As$_2$ near the Γ point [$\mathbf{k} = (0,0,0)$] by a single 3D-Dirac cone, $\epsilon_{\mathbf{k}} = \hbar v_F |\mathbf{k}|$. Here, $k_B$ denotes the Boltzmann constant, and we have set the Fermi velocity $v_F=0.93\times 10^6$ m/s[26], the Fermi energy $E_F=50$ meV, and the temperature $T=300$ K. For simplicity, we have assumed that neither $f_{FD}(\mathbf{k})$ nor $\tau_s$ depends on time even when intense ac field $\mathbf{E}(t)$ is applied. Our experimental results in Figs. 2(g) and 2(h) demonstrate that the dynamics of the electron density and the scattering time $\tau_s$ are much slower than THz field cycle and do not change largely, indicating the validity of this assumption.

We adopt a linearly-polarized $\boldsymbol{E}(t)$ with the Gaussian envelope. Owing to the isotropy, we assume that $\boldsymbol{E}(t)$ is along the $x$-axis as $\boldsymbol{E}(t) = (E_x(t), 0, 0)$ without loss of generality:

$$E_x(t) = E_0 \cos(2\pi f t) \exp\left(-\ln 2 \left(\frac{2t}{t_{FWHM}}\right)^2\right). \quad (5)$$

Here $E_0$ is the peak amplitude, $f$ the center frequency, and $t_{FWHM}$ the full width at half maximum of the electric-field envelope. Referring to our THz field, we have set $t_{FWHM}$=8 ps. We assume that $E_x(t)$ is uniform in space, which implies that the wavelength of our THz wave is sufficiently larger than typical length scales of the system and the sample dimensions.

We numerically solve the time evolution Eq. (4) for our pulse electric field (4) by discretizing $\boldsymbol{k}$ and $t$. We take the initial time of calculation $t_{\text{init}}$ ($< 0$) so small that $\boldsymbol{E}(t_{\text{init}}) \sim 0$, and set $\rho(\boldsymbol{k}, t_{\text{init}}) = f_{FD}(\boldsymbol{k})$. Noticing that the equation is symmetric under rotations in the $k_y k_z$ plane, we can characterize the electron density as $\rho(\boldsymbol{k}, t) = \rho'(k_x, k_\perp, t)$ with $k_\perp = \sqrt{k_y^2 + k_z^2}$.

The time-dependent electric current density $\boldsymbol{J}(t)$ is given by

$$\boldsymbol{J}(t) = 4e v_F \int \frac{d\boldsymbol{k}}{(2\pi)^3} \rho(\boldsymbol{k}, t) \hat{\boldsymbol{k}},$$

where $\hat{\boldsymbol{k}} = \boldsymbol{k}/|\boldsymbol{k}|$ and the factor of 4 derives from the four-fold degeneracy of the band including the spin degree of freedom. The $\boldsymbol{k}$-integral is taken over the Brillouin zone, but can be extended to the entire $\boldsymbol{k}$-space because $\rho(\boldsymbol{k}, t)$ is nonvanishing only in the vicinity of the Γ point in our setup. By invoking the rotational invariance in the $k_y k_z$ plane, we have $J_y(t) = J_z(t) = 0$ and

$$J_x(t) = \frac{e v_F}{\pi^2} \int_{-\infty}^{\infty} dk_x \int_0^{\infty} dk_\perp k_\perp \rho'(k_x, k_\perp, t) \frac{k_x}{\sqrt{k_x^2 + k_\perp^2}},$$

which is readily obtained from the solution of the time evolution equation.

The experimentally-observed transmitted field $\boldsymbol{E}_{trans}(t)$ is obtained by [S13]

$$\boldsymbol{E}_{trans}(t) = \frac{2\boldsymbol{E}(t) - Z_0 d \boldsymbol{J}(t)}{n_s + 1} \frac{2 n_s}{1 + n_s}, \quad (6)$$

where $Z_0$ is the vacuum impedance, $d$ the sample thickness, and $n_s$ the refraction index of the substrate. We have confirmed that the THz wavelength is much longer than the sample thickness and the effect of electric-field propagation inside the sample is negligibly small. Fourier transforming Eq. (6), we obtain the spectra for the transmitted amplitude $|\tilde{\boldsymbol{E}}_{trans}(\omega)|$ and intensity $\propto |\tilde{\boldsymbol{E}}_{trans}(\omega)|^2$. We note that the Fourier transform of the incident field $\boldsymbol{E}(t)$ is peaked at the center frequency and does not have significant amplitude at higher harmonics. Thus, the harmonic peaks of, say, $|\tilde{\boldsymbol{E}}_{trans}(\omega)|$ derive from those of $\boldsymbol{J}(t)$, which is induced by the strong coherent driving of electrons within the band.

## 11. Calculation in the thermodynamic model

Following a previous literature [S14], we performed the simulation of the thermodynamic model for Cd$_3$As$_2$ thin film.

The key concept for the nonlinearity in the thermodynamic model [S14] is the change of the Drude-type response function including the scattering time $\tau_s$, which is enhanced with heating as $\tau_s(\Delta Q) = \tau_s^0 + \gamma \Delta Q/N_c$. Here $\tau_s^0$=145 fs is the scattering time at room temperature, $N_c$ is the carrier density per area, $\Delta Q$ is the heat transfer per area from a THz pulse to a sample, and $\gamma$ is the proportionality constant expressed as $\gamma = \tau_s^0/E_F$ ($E_F$=50 meV and $N_c$=2.7×10$^{12}$ cm$^{-2}$ are the same as in our intraband acceleration model). Such an increase of $\tau_s$ by heating seems to be counterintuitive in usual materials but might occur in a Dirac system as theoretically studied in term of the Coulomb screening [S4-S6]. Actually, a previous nonlinear THz transmission experiment for graphene [S15] has been interpreted by the enhancement of $\tau_s$. For our Cd$_3$As$_2$ film sample, the enhancement of $\tau_s$ is observed with increase of temperature in Fig. S1(f) and with THz pumping in Fig. 2(h), which satisfy this assumption of the thermodynamic model.

Figure S8(a) shows the calculated amplitude spectra for the relaxation time $\tau_R$=0.5 and 8 ps with the field strength $E_{pump}$=31 kV/cm, which indicates that the HHG amplitude becomes weaker for longer relaxation. The effect of $\tau_R$ on the nonlinearity response is more clearly seen in Fig. S8b, where the THG amplitude is plotted as a function of $E_{pump}$ for various $\tau_R$ in comparison with the experimental data. For longer relaxation time, the saturation occurs in weaker excitation regime. It can be accounted for by the effect of heating; when the relaxation time is longer than the THz field cycle, the effect of the heat transfer is piled up for every field cycle and then the response easily goes into the nonperturbative regime. In the case of $\tau_R$= 8 ps, the THG amplitude remains around 10 V/cm for $E_{pump}$=31 kV/cm, which is an order of magnitude smaller than the experimental result. Both of the THG amplitude and the saturating behavior are not reproduced by the thermodynamic model with $\tau_R$= 8 ps.

Moreover, another crucial inconsistency is also discerned in Fig. S8(c) which shows the temporal evolution of the scattering time $\tau_s$ for various $\tau_R$. During the pump THz field irradiation, $\tau_s$ oscillates in time as a source of the THG. When $\tau_R$ is longer than the field cycle, the heating occurs significantly and results in an anomalous enhancement of $\tau_s$ up to ~1200 fs. On the other hand, the change of $\tau_s$ was experimentally observed by THz pump-THz probe spectroscopy in Fig. 2(h). The enhancement of $\tau_s$ is from 145 to 155 fs even for the strong pumping of 50 kV/cm. This result means that $\tau_s$ can increase only 7 % even for such a strong field, which does not agree with the significant enhancement of $\tau_s$ in the thermodynamic model.

We have also run the simulation with a smaller proportionality constant $\gamma$, for which the effect of heating is reduced and therefore the enhancement of $\tau_s$ can be suppressed in the thermodynamic model. In this case, however, the THG amplitude becomes even much smaller.

Therefore, we conclude that the thermodynamic model cannot reproduce the efficient THG with a long relaxation time.

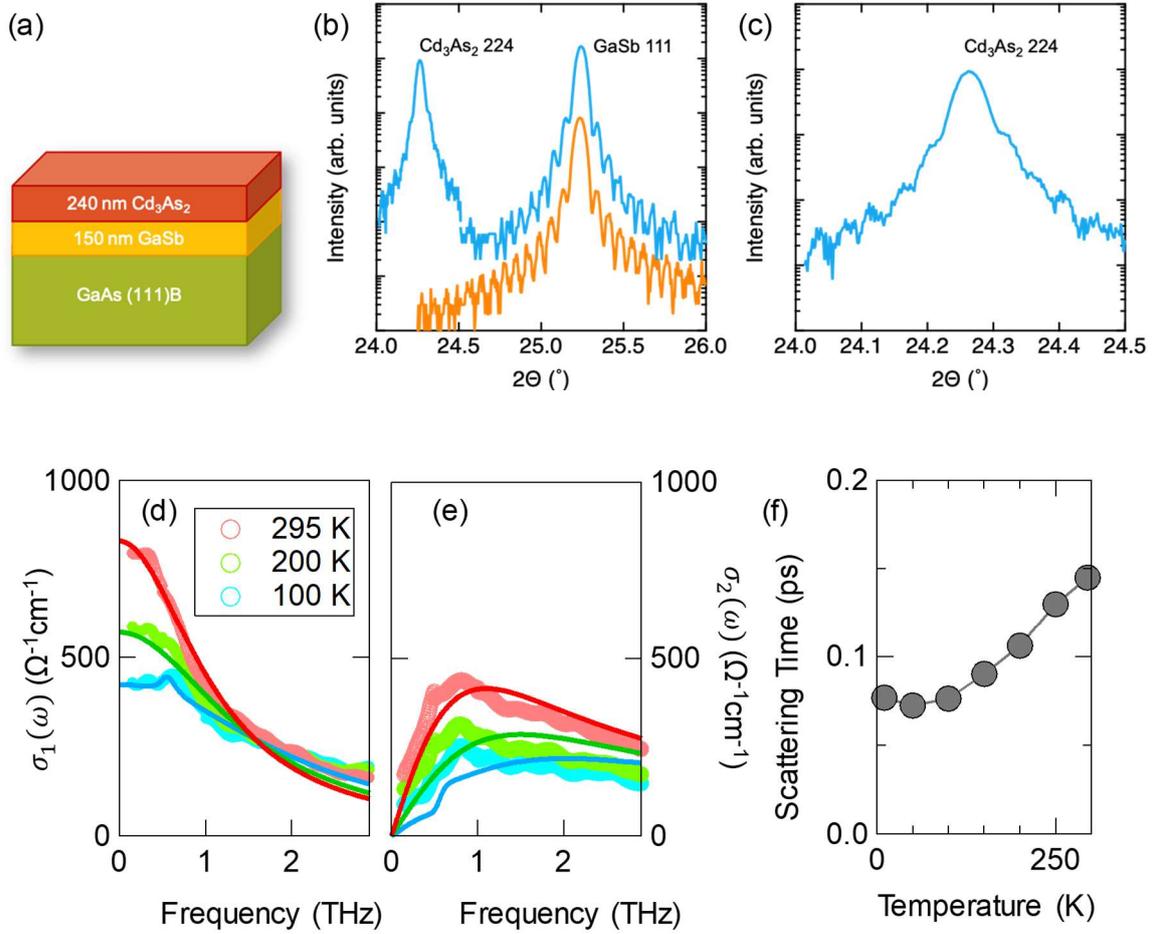

**Figure S1.** (a) Structure of the sample. (b) X-ray diffraction pattern of the GaSb 111 and Cd$_3$As$_2$ 224 reflection. The blue curve was obtained from the Cd$_3$As$_2$/GaSb/GaAs sample and the orange one was obtained from the GaSb/GaAs reference sample. (c) High-resolution diffraction pattern of the Cd$_3$As$_2$ 224 reflection. (d)(e) Real- and imaginary-part THz conductivity spectrum at various temperature. The solid curves are the results of fitting. (f) Temperature dependence of the scattering time obtained by the fitting.

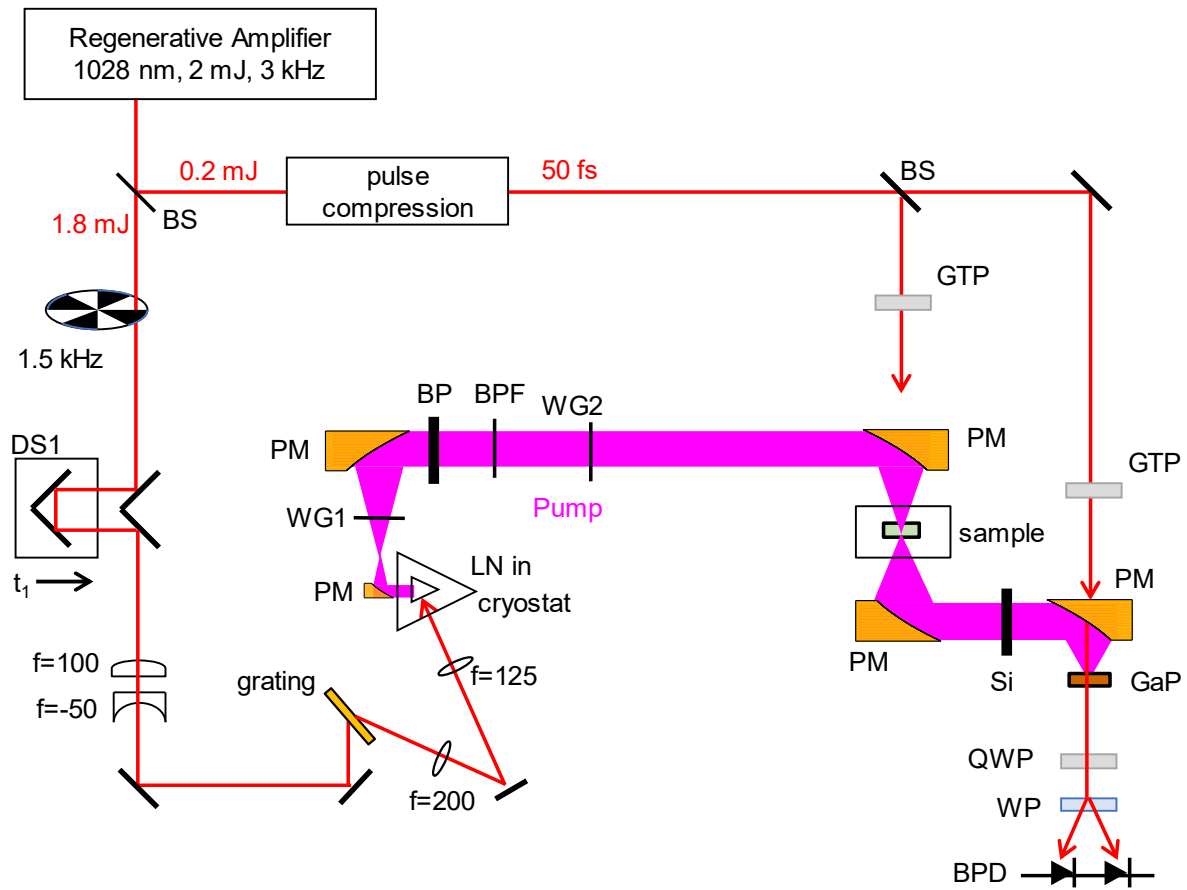

**Figure S2.** BS: beam splitter, DS: delay stage, LN: LiNbO$_3$ crystal, PM: parabolic mirror, WG: wire-grid polarizer, BP: black polyethylene, BPF: metal-mesh band-pass filters, GTP: Glan–Thompson prism, QWP: quarter-half waveplate, WP: Wollaston prism, BPD: balanced photodiode. The inset shows the calculated detection efficiency of the EO sampling in GaP crystals.

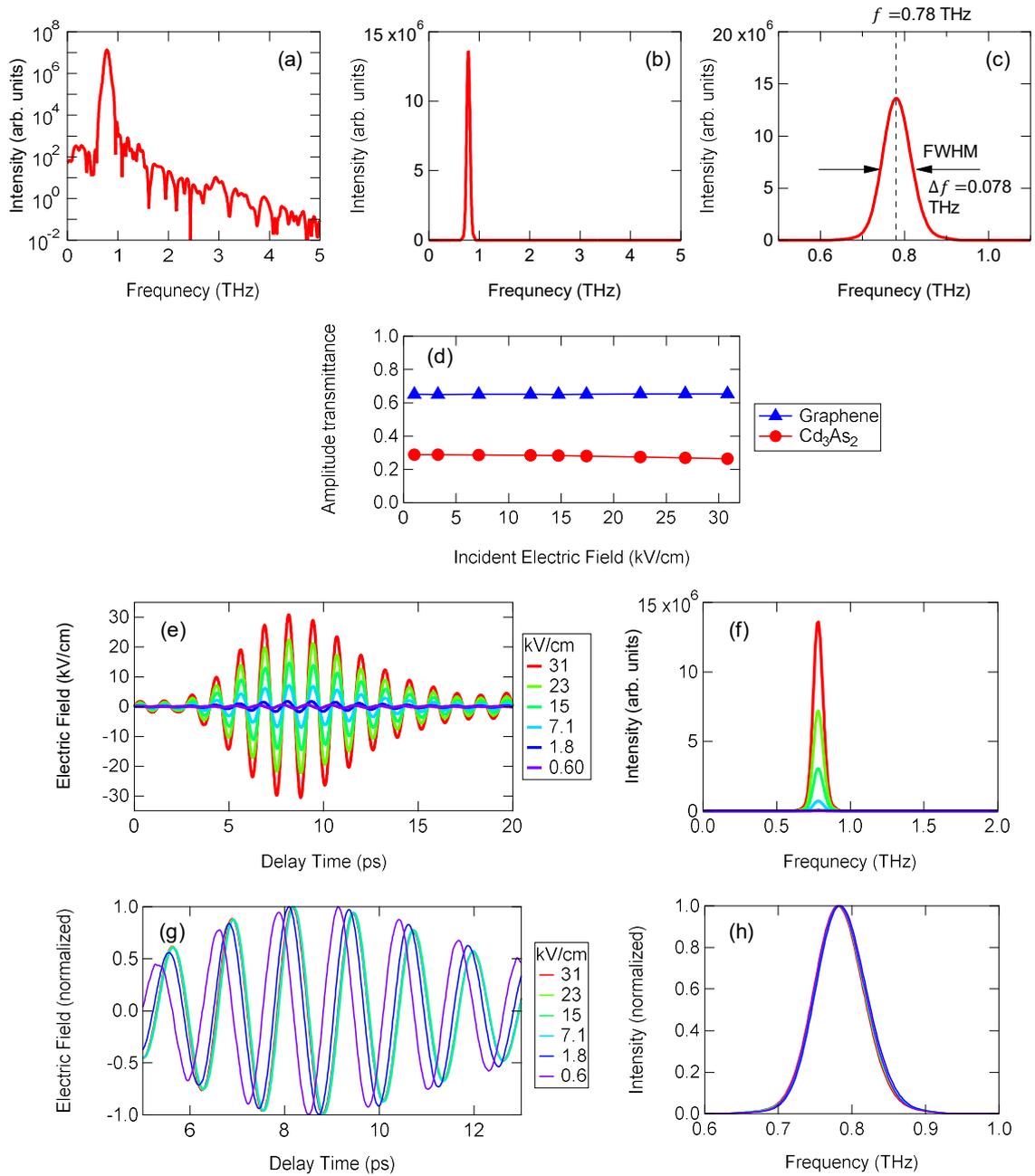

**Figure S3.** (a)-(c) The power spectra of the THz pump pulse used for THG experiment with logarithm scale, linear scale, and extended linear scale, respectively. (d) Transmittance of the fundamental component as a function of incident electric field. (e)(f) The waveforms and power spectra of the THz pump pulse for various field strengths controlled by the wire-grid polarizer. (g)(h) The same data with e and f normalized for comparison.

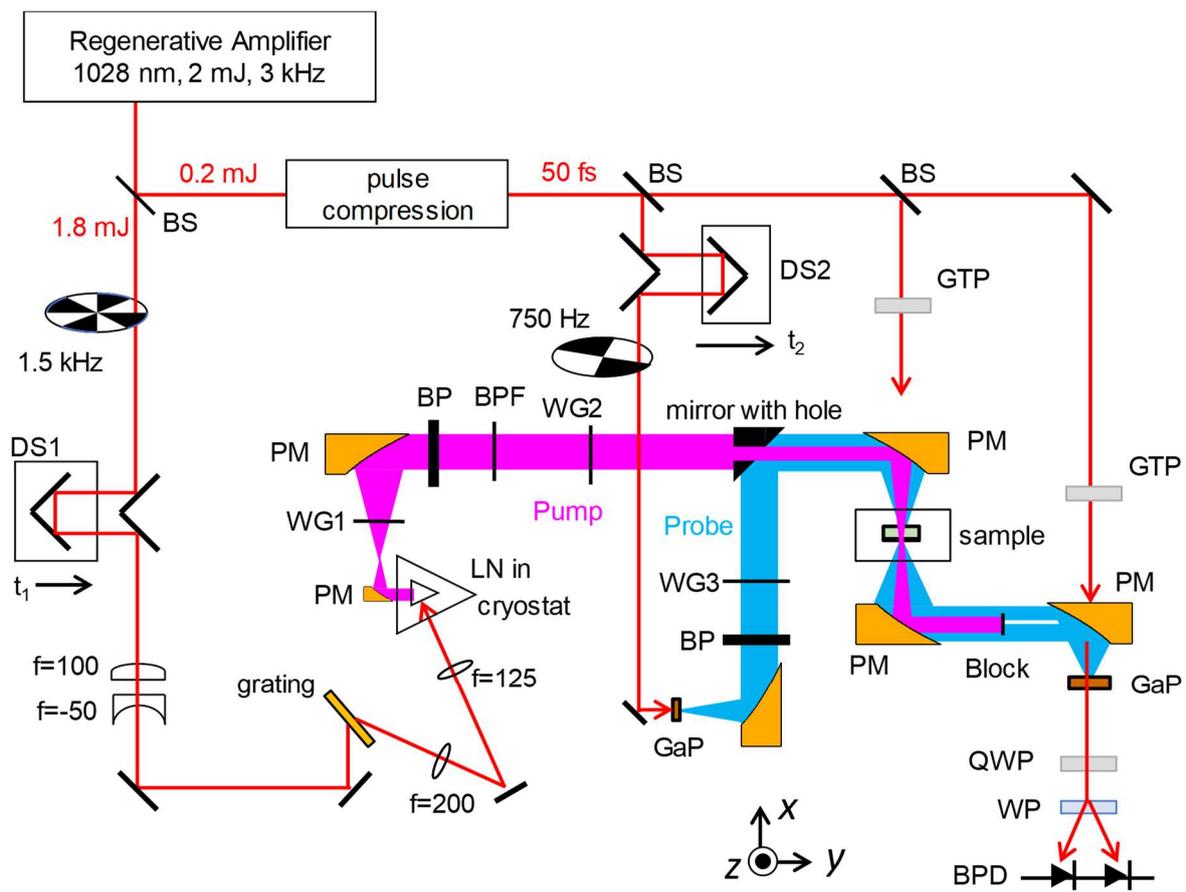
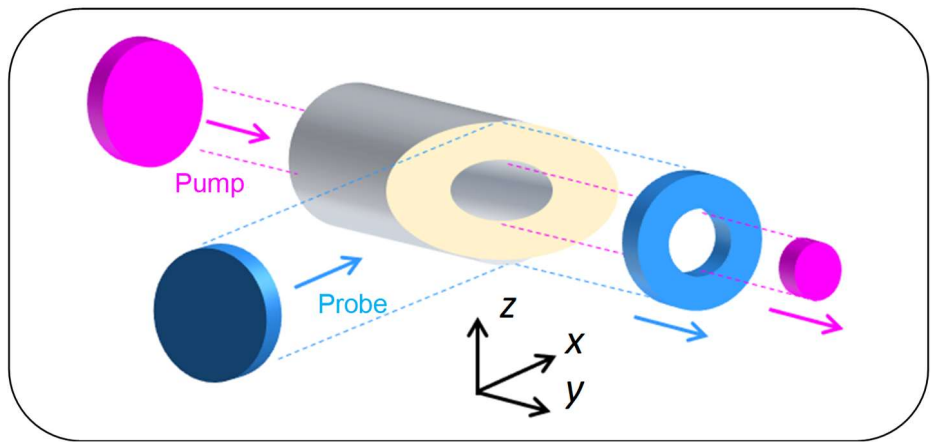

**Figure S4.** The inset shows a mirror combining the THz pump pulses and THz probe pulses.

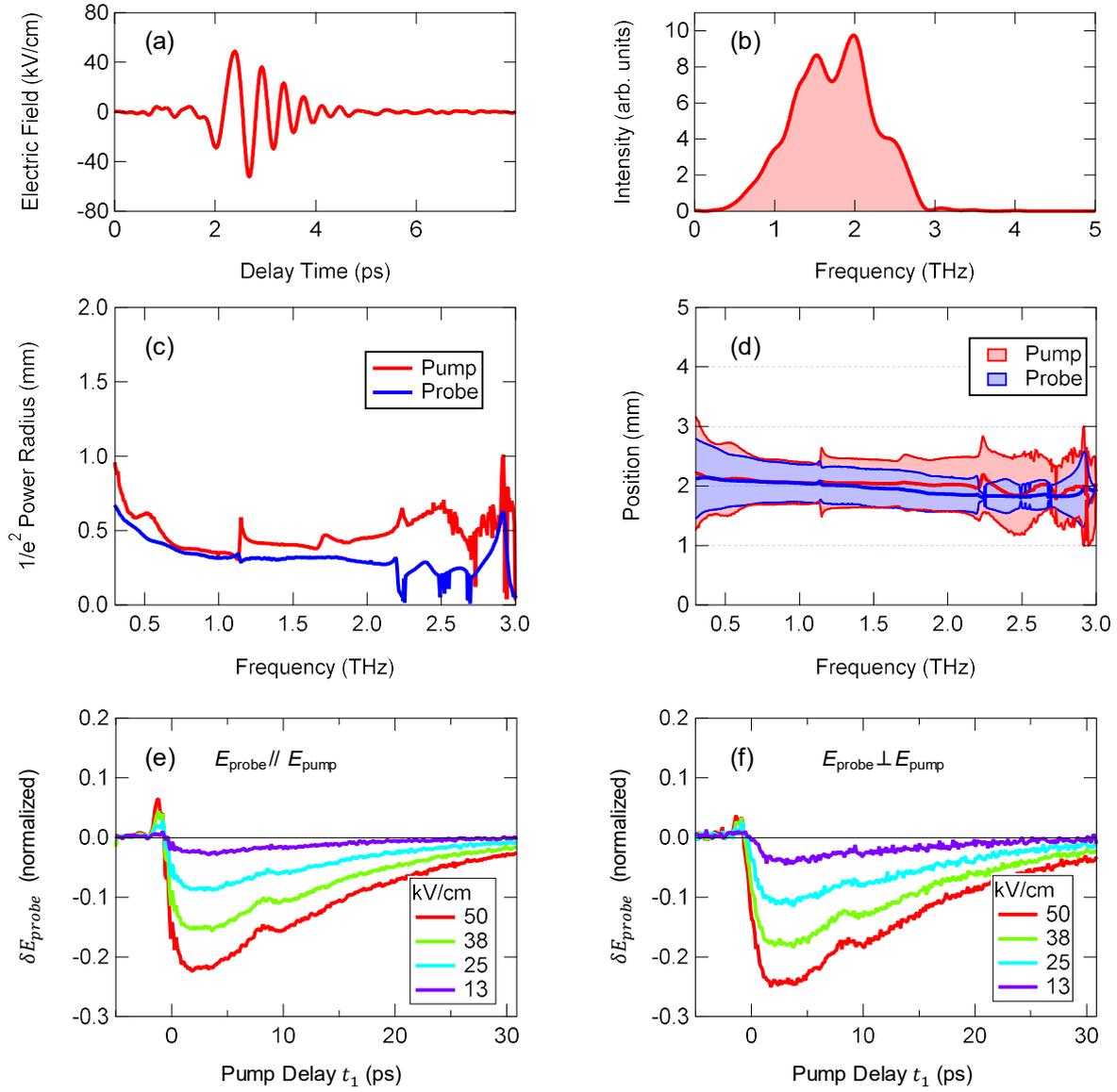

**Figure S5.** (a)(b) The waveform and power spectrum of the broad band short pump pulse. (c) The radius of the pump and probe pulse at the sample position evaluated by the knife-edge method. (d) The spatial overlap of the pump and probe pulses at the position of the sample. The solid curves show the center position of the pulses. The colored areas indicate the beam diameters. (e)(f) The pump-probe dynamics $\delta E_{probe}$ as a function of $t_1$ for various pump field strengths with the probe polarization parallel and perpendicular to the pump.

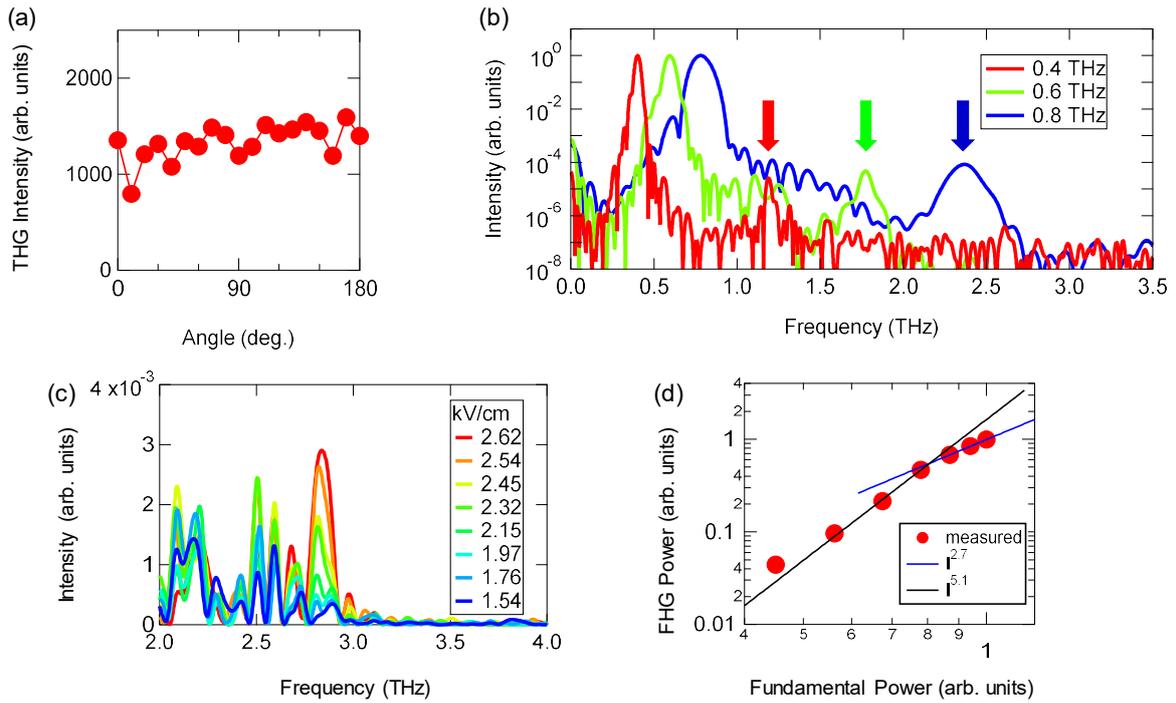

**Figure S6.** (a) Angular dependence of the THG intensity. (b) Transmitted THz spectra for 280-nm sample with different excitation frequencies of $f$=0.4, 0.6, and 0.8 THz. The THG signal at $3f$ are indicated by arrows. (c) The spectra of FHG at the excitation frequency of $f$=0.6 THz for the 240-nm sample. (d) The FHG intensity as a function of the fundamental power.

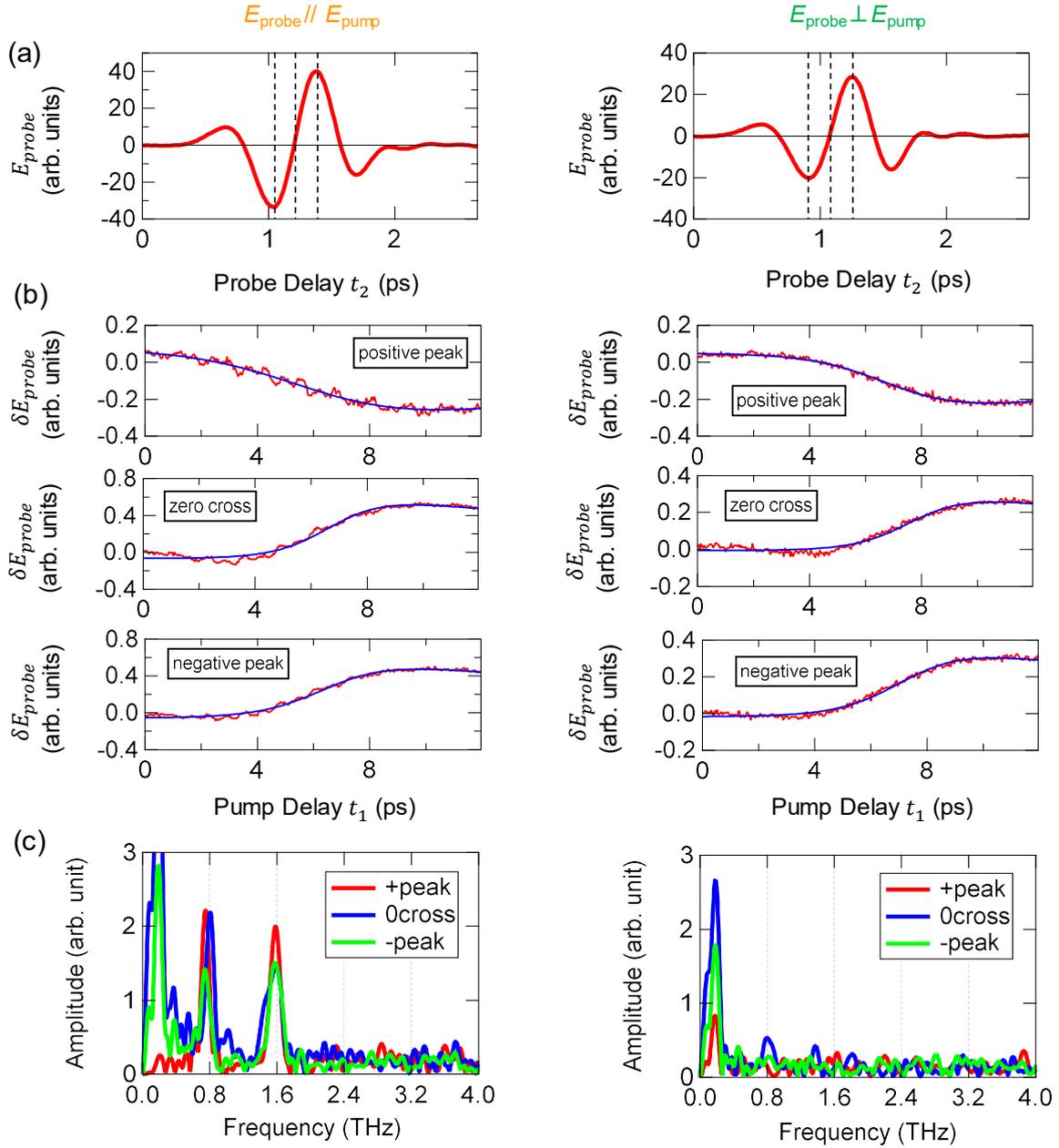

**Figure S7.** (a) Probe pulse waveform. The dotted lines indicate the fixed probe delay point for the pump-probe measurement. (b) The pump-probe signal during the quasi-monochromatic pump excitation with 0.8-THz frequency at each fixed probe delay point. (c) Fourier components in the pump-probe signals. The left and right columns correspond to the data for $E_{probe} \parallel E_{pump}$ and $E_{probe} \perp E_{pump}$, respectively.

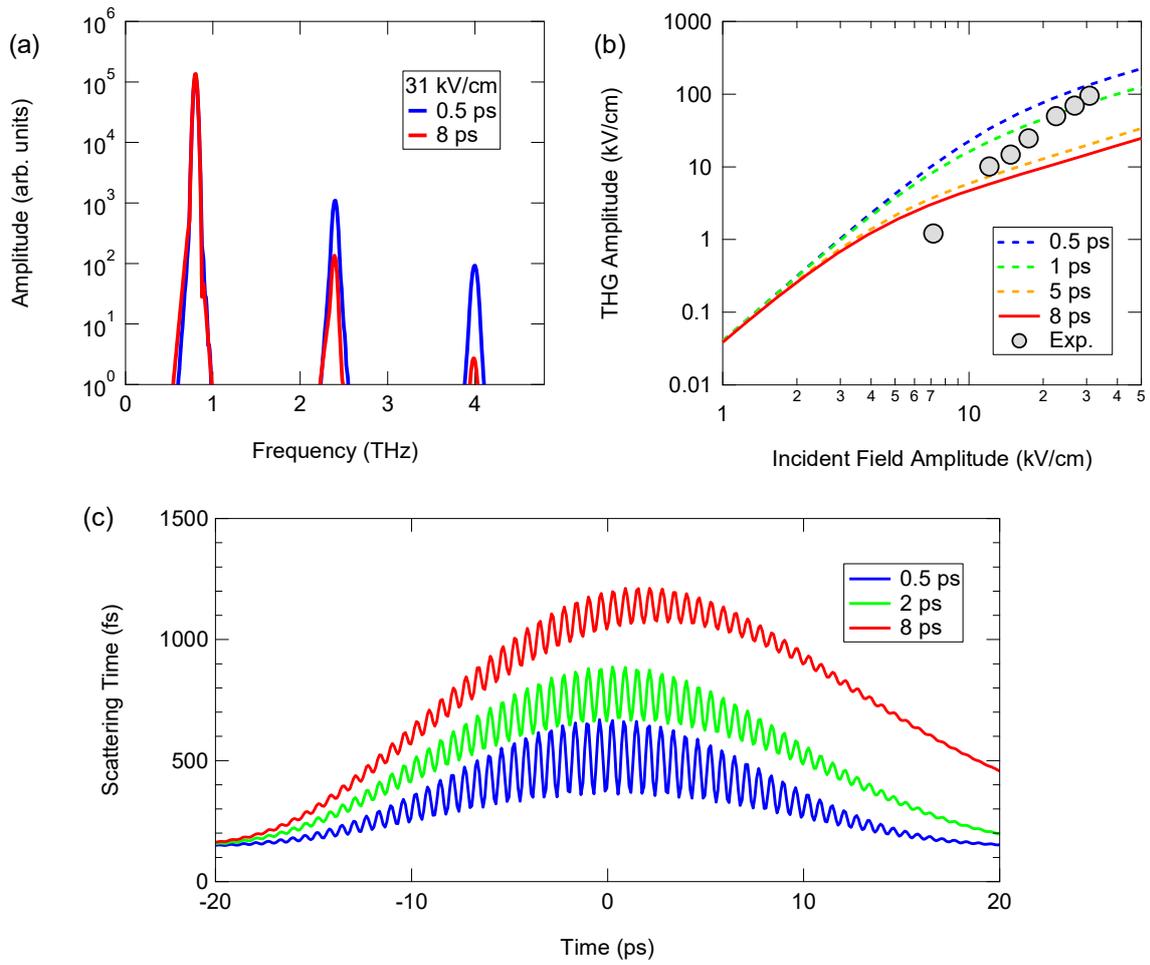

**Figure S8.** (a) Calculated amplitude spectra for $E_{pump}$=31 kV/cm for the relaxation time $\tau_R$=0.5 and 8 ps. (b) THG amplitude as a function of $E_{pump}$ for various $\tau_R$. (c) Temporal change of the scattering time $\tau_s$ for various $\tau_R$.